\documentstyle[12pt,epsfig]{article}

\newcommand{\be}{\begin{equation}}
\newcommand{\ee}{\end{equation}}
\newcommand{\bea}{\begin{eqnarray}}
\newcommand{\eea}{\end{eqnarray}}

\newcommand{\ms}{\Delta m^2_{21}}
\newcommand{\ma}{\Delta m^2_{\rm atm}}

\newcommand{\sss}{\sin^2 \theta_{12}}
\newcommand{\sch}{\sin^2 \theta_{13}}

\newcommand{\kl}{\mbox{KamLAND~}}
\newcommand{\beq}{\begin{equation}}
\newcommand{\eeq}{\end{equation}}

\def\nue{{\nu_e}}
\def\anue{{\bar\nu_e}}

\def\gtap{\ \raisebox{-.4ex}{\rlap{$\sim$}} \raisebox{.4ex}{$>$}\ }

\def\gtap{\ \raisebox{-.4ex}{\rlap{$\sim$}} \raisebox{.4ex}{$>$}\ }

\begin{document}

\begin{flushright}
SISSA 26/2004/EP\\
hep-ph/0404103
\end{flushright}

\begin{center}
{\Large \bf 
Reactor Anti-Neutrino Oscillations and Gadolinium Loaded 
Super-Kamiokande Detector}
\vspace{.5in}

Sandhya Choubey$^{1,2}$
and
S.T. Petcov$^{2,1,3,4}$\vskip .5cm
$^1$INFN, Sezione di Trieste, Trieste, Italy.\\
$^2$Scuola Internazionale Superiore di Studi Avanzati, 
I-34014 Trieste, Italy.\\
$^3$Yukawa Institute for Theoretical Physics, 
Kyoto University, Kyoto 606-8502, Japan.\\
$^4$Institute of Nuclear Research and
Nuclear Energy, Bulgarian Academy of Sciences, 1784 Sofia, Bulgaria.\\
\vskip 1in

\end{center}

\begin{abstract}
\vskip 1cm
We explore the potential of measuring the solar neutrino oscillation 
parameters in the proposed gadolinium loaded Super-Kamiokande (SK-Gd) 
detector. Gadolinium dissolved in water can detect neutrons 
much more efficiently than pure water. This imparts the
detector the ability to observe electron type antineutrinos, 
transforming Super-Kamiokande 
into a huge reactor antineutrino detector with 
an event rate approximately 43 times higher than 
that observed in KamLAND. We simulate the 
reactor antineutrino 
data expected in this high statistics detector. 
We use these prospective data to study the 
precision with which the solar neutrino 
oscillation parameters,
$\Delta m^2_{\odot}$ and $\sin^2\theta_{\odot}$,
can be determined 
i) with the SK-Gd detector, and 
ii) by combining the SK-Gd data with the global data
on solar neutrino oscillations.
For comparison and completeness
the allowed regions of 
$\Delta m^2_{\odot}$ and $\sin^2\theta_{\odot}$,
expected to be obtained from the data of the solar neutrino 
and KamLAND experiments, are also presented. 
We find that the SK-Gd experiment could provide one of the most
precise (if not the most precise) determination of the
solar neutrino oscillation parameters
$\Delta m^2_{\odot}$ and $\sin^2\theta_{\odot}$.
\end{abstract}





\newpage

\section{Introduction}
\label{section:introduction}

  At present we have compelling evidences for
oscillations of both solar ($\nu_e$) and
atmospheric ($\nu_{\mu}$ and $\bar{\nu}_{\mu}$)
neutrinos, driven by nonzero neutrino masses 
and neutrino mixing \cite{BPont57,Pont67}.
They have been accumulated over a long period of time
in the experiments
with solar neutrinos
\cite{Cl98,Kamiokande,SAGE,GALLGNO,SKsol,SNO1,SNO2,SNO3}, 
Homestake, Kamiokande, SAGE, GALLEX/GNO, Super-Kamiokande and SNO,
in the KamLAND experiment with 
reactor antineutrinos \cite{KamLAND},
and in the studies of the fluxes of 
atmospheric neutrinos by the 
Super-Kamiokande collaboration 
\cite{SKatmo98,SKatmo03,SKdip04}.
Indications for oscillations of neutrinos have 
also been obtained in the K2K long base-line
experiment \cite{K2K}.

  Evidences for oscillations of the solar  
$\nu_e$, induced by nonzero neutrino masses 
and neutrino mixing \cite{Pont67}, 
have been reported first by 
the pioneering Davis et al. (Homestake) 
experiment \cite{Pont46,Davis68}. They have 
been confirmed and reinforced 
later by Kamiokande, SAGE, GALLEX/GNO 
and Super-Kamiokande experiments.
The evidences for mixing and
oscillations of the solar $\nu_e$ have been 
made compelling during the last three years
by the data from the SNO solar neutrino 
and KamLAND reactor antineutrino experiments
\cite{SNO1,SNO2,SNO3,KamLAND} 
(see also, e.g., \cite{snonc}).
Under the assumption of CPT-invariance,
the observed disappearance of 
reactor $\bar{\nu}_e$ 
in the KamLAND experiment, in particular,
confirmed the interpretation 
of the solar neutrino data
in terms of $\nu_e \rightarrow \nu_{\mu,\tau}$
oscillations, induced by nonzero 
neutrino masses and 
nontrivial neutrino mixing.
The KamLAND results \cite{KamLAND}
practically established the
large mixing angle (LMA)
MSW solution (see, e.g., \cite{SNO1}) 
as unique solution of the solar neutrino problem.
The solar neutrino and KamLAND data, including 
the salt phase SNO results, strongly favor the 
low-LMA solution with 
\cite{SNO3,salt,saltother}
$\Delta m^2_{21} \sim 7\times 10^{-5}$ 
eV$^2$ and $\sin^2\theta_{12} \sim 0.3$,
$\ms \equiv \Delta m^2_{\odot}$ and
$\theta_{12} \equiv \theta_{\odot}$
being the neutrino mass squared difference and mixing
angle driving the solar neutrino oscillations.
The high-LMA solution (see, e.g., \cite{KamLAND}),
characterized by $\Delta m^2_{21} \sim 1.5\times 10^{-4}$ eV$^2$ 
and similar value of $\sin^2\theta_{12}$, is only allowed at 
99.14\% C.L. by the data \cite{salt}.

  Strong evidences for
oscillations of the atmospheric 
$\nu_{\mu}$ ($\bar{\nu}_{\mu}$) 
have been obtained in the Super-Kamiokande
experiment from the observed i) Zenith angle dependence 
of the multi-GeV and sub-GeV 
$\mu$-like events \cite{SKatmo98,SKatmo03}, 
and ii) the recently reported observation
of an ``oscillation dip'' in the $L/E-$dependence
of the (essentially multi-GeV) 
$\mu-$like atmospheric neutrino events \cite{SKdip04}
\footnote{The sample used in the analysis of the
$L/E$ dependence consists of 
$\mu-$like events 
for which the relative uncertainty in the 
experimental determination of the 
$L/E$ ratio does not exceed 70\%.},
$L$ and $E$ being the 
distance traveled by neutrinos and 
the neutrino energy.
As is well known, the SK atmospheric 
neutrino data is best described in 
terms of dominant two-neutrino
$\nu_{\mu} \rightarrow \nu_{\tau}$ 
($\bar{\nu}_{\mu} \rightarrow \bar{\nu}_{\tau}$)
vacuum oscillations  with \cite{SKatmo03,SKdip04}
$\ma \equiv \Delta m^2_{31} \sim (2.0 - 3.0)\times 10^{-3}$ eV$^2$ 
and maximal mixing,
$\sin^22\theta_{\rm atm} \equiv \sin^22\theta_{23}\sim 1.0$.
The observed dip is predicted 
due to the oscillatory dependence on $L/E$
of the $\nu_{\mu} \rightarrow \nu_{\tau}$ and
$\bar{\nu}_{\mu} \rightarrow \bar{\nu}_{\tau}$
oscillation probabilities. 
This beautiful result represents the first 
ever observation of a direct effect 
of the oscillatory dependence 
on $L/E$ of the probability 
of  neutrino oscillations in vacuum.

  Information on the third mixing angle
$\theta_{13}$, present   
in the case of 3-neutrino mixing which is
required to describe the solar and
atmospheric neutrino (and KamLAND)
data in terms of neutrino oscillations,
is provided essentially by the 
short baseline reactor antineutrino experiments 
CHOOZ and Palo Verde \cite{CHOOZPV}.
Using the range of allowed values of $\ma$ 
found in the latest SK data analysis 
\cite{SKatmo03,Fogliatm0308055}, 
a combined 3-$\nu$
oscillation analysis of the solar neutrino,
CHOOZ and KamLAND data gives 
(at 99.73\% C.L.)
$\sin^2 \theta_{13} < 0.074$ \cite{salt}.
A somewhat more stringent limit 
was obtained in the global analysis of all the available 
solar, atmospheric and reactor neutrino data,
performed in \cite{ConchaNOON04}:
$\sin^2 \theta_{13} < 0.050$ (99.73\% C.L.).
 
  Neutrino flavor oscillations have also been claimed to have 
been observed in the LSND experiment \cite{lsnd}. This interpretation
of the LSND results is being currently tested in the 
MiniBOONE experiment \cite{mboone}.

  After the remarkable progress made in the last few years 
in establishing the existence of neutrino oscillations,
one of the main goals of the future experimental 
studies of neutrino mixing is to measure with high precision
the parameters which drive the solar and atmospheric neutrino
oscillations, $\Delta m^2_{21}$, $\Delta m^2_{31}$,
$\sin^2\theta_{12}$ and $\sin^2\theta_{23}$.
The potential of the current and future experiments 
for high precision determination 
of the solar neutrino oscillation parameters,
$\Delta m^2_{21}$ and $\sin^2\theta_{12}$, 
was studied recently in 
\cite{th12,th12jnb,th12hlma,shika} 
(see also \cite{talks}).
The precision in the measurement of 
$\Delta m^2_{21}$ and $\sin^2\theta_{12}$,
which can be  reached
in the Borexino and the so-called 
LowNu experiments, was explored in \cite{th12,th12jnb}.
The impact of the prospective 
increase in statistics of the \kl data 
on the determination of 
$\Delta m^2_{21}$ and $\sin^2\theta_{12}$
was investigated in detail in \cite{salt,th12,shika}. 
If the $e^+-spectrum$ measured at KamLAND
is simulated at a point in the low-LMA region, 
the allowed 3$\sigma$ area in the 
high-LMA zone reduces in size 
in the case of 0.41 kTy of data, and disappears if
the statistics is increased to 1.0 kTy \cite{salt}.
In this case $\Delta m^2_{21}$ will be 
determined with high precision.
If, however, the spectrum observed in \kl  
conforms to a point in the high-LMA region,
the conflicting trend of solar and 
\kl data would make the high-LMA
solution reappear at 90\% C.L. 
and the determination of $\Delta m^2_{21}$ would
remain ambiguous \cite{salt}. In both cases
$\sin^2\theta_{12}$ cannot be determined 
with high precision. Actually,
reaching high accuracy in the measurement
of $\sin^2\theta_{12}$ (e.g., 10\% error 
at 1$\sigma$) is a rather challenging
problem. It can be solved by performing a 
dedicated reactor $\bar{\nu}_e$ 
experiment \cite{th12,th12hlma}, or 
by the future solar neutrino experiments
LENS and XMASS, aiming to measure the pp neutrino flux 
\cite{LENSNOON04,XMASSNOON04}.

  Recently it was proposed in \cite{gdpaper} 
to dope the SK detector with Gadolinium 
by dissolving 0.2\% gadolinium trichloride in the water
\footnote{The authors of \cite{gdpaper}  
called  ``GADZOOKS!'' the resulting detector.
We will use in what follows the abbreviation SK-Gd for it.}.
The added gadolinium would make it possible
to detect the neutrons, released 
in the $\anue$ capture on protons, with a 
relatively high efficiency. 
This would allow the SK experiment, 
in particular, to detect the reactor antineutrinos 
coming from the numerous powerful nuclear 
reactors located in Japan,
thus transforming SK
into a huge reactor 
antineutrino detector with 
an event rate approximately 43 times higher than 
that observed in KamLAND.
In this paper we study the 
prospects of measuring the solar 
neutrino oscillation parameters by 
observing the reactor  antineutrino 
oscillations in the proposed 
gadolinium loaded SK (SK-Gd) experiment. 
Since $\Delta m^2_{21} \ll \Delta m^2_{31}$
and the mixing angle $\theta_{13}$ is
restricted to be 
relatively small, the
third (heaviest) neutrino with definite mass
is expected to have  negligible impact 
on the determination of the values of 
the solar neutrino oscillation parameters
\footnote{The effect of  $\sch$
can become important if 
$\sin^22\theta_{12}$ is measured
with a $\sim 10\%$ precision,
which, as we will see, is higher 
than the precision which can be
reached in the SK-Gd experiment.}
in the experiment of interest 
and we perform our study in the framework of 
the 2$\nu-$ mixing scenario ($\sch = 0$).

  We begin in Section \ref{solklsec}
with an overview of the 
currently allowed ranges of values of the
solar neutrino oscillation parameters 
by the existing global solar 
neutrino and KamLAND data. 
We present further the
regions in the 
$\Delta m^2_{21} - \sin^2\theta_{12}$ plane,
which are expected to be allowed 
after taking into account 
future higher statistics and 
lower systematic data from SNO and from KamLAND
experiments. The SK-Gd detector
is considered  in Section \ref{sens1},
where we discuss the assumptions 
made to simulate the data in this 
proposed modification of the SK experiment, 
as well as
the procedure used to statistically
analyze the simulated 
data. We investigate further
the potential of the SK-Gd experiment in reducing the 
uncertainties in the values of 
$\ms$ and $\sss$.
In Section \ref{solgdsec} the expected results 
from a combined analysis of the solar neutrino data and the
prospective SK-Gd simulated spectrum data are given. 
In Sections \ref{sens1} and \ref{solgdsec} 
we also discuss the bounds on 
$\ms$ and $\sss$ 
one would obtain if some 
of the reactors in Japan 
would be ``switched off''. Section \ref{concl} 
contains the main conclusions of our study.

\section{Solar Neutrino Oscillation Parameters 
from the Solar Neutrino 
and KamLAND Data}
\label{solklsec}

\begin{figure}[t]
\centerline{\psfig{figure=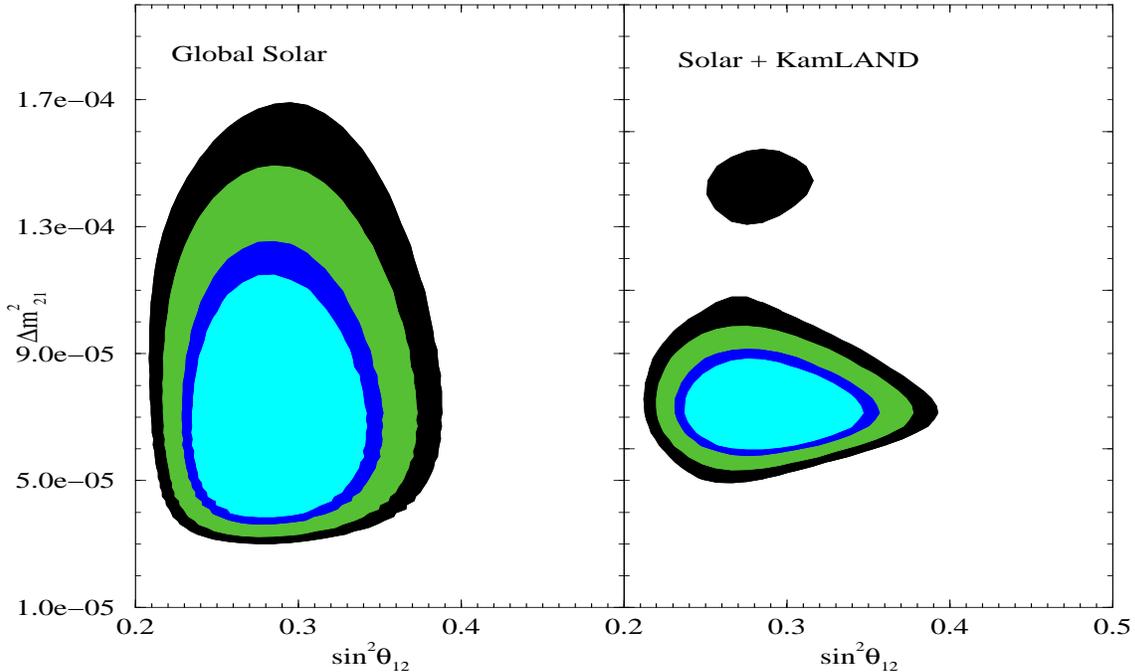,height=9cm,width=15cm}}
\caption{\label{solnow}
The 90\%, 95\%, 99\%, 99.73\% C.L. 
allowed regions in the $\ms-\sss$ plane. 
The left-hand and right-hand panels show the allowed regions 
obtained in $\chi^2-$analysis 
respectively of the global data from 
the solar neutrino experiments
and of the solar neutrino and KamLAND data. 
}
\end{figure}

\begin{table}[p]
\begin{center}
\begin{tabular}{ccccc}
\hline
{Data} & 99\% CL &99\% CL  & 99\% CL
& 99\% CL \cr
{set} & range of & spread &range  & spread
\cr
{used} & $\Delta m^2_{21}\times$ & of &
of & in  \cr
& 10$^{-5}$eV$^2$ & {$\Delta m^2_{21}$} & $\sin^2\theta_{12}$
& {$\sin^2\theta_{12}$} \cr
\hline
{only solar} & 3.2 - 14.9
&{65\%} & $0.22-0.37$ &{25\%}\cr
{solar+162 Ty KL}&  5.2 - 9.8
& {31\%}
& $0.22-0.37$ &{25\%}  \cr
solar with future SNO & $3.3-11.9$ & 57\% & $0.22-0.34$ & 21\% \cr
{solar+1 kTy KL(low-LMA)}& 6.5 - 8.0
& {10\%} &
$0.23-0.37$ & {23\%}\cr
solar+2.6 kTy KL(low-LMA) & $6.7-7.7$ & 7\% & $0.23-0.36$ & 22\% \cr 
solar with future SNO+1.3 kTy KL(low-LMA) & $6.7-7.8$ & 8\% & $0.24-0.34$ & 17\% \cr 
3 yrs SK-Gd & 7.0 - 7.4 & 3\% & $0.25-0.37$ & 19\% \cr
5 yrs SK-Gd & $7.0-7.3$ & $2\%$ & $0.26-0.35$ & 15\% \cr
solar+3 yrs SK-Gd(low-LMA) & $ 7.0-7.4$ & 3\% & $0.25-0.34$ & 15\%\cr
solar+3 yrs SK-Gd(high-LMA) & $ 14.5-15.4$ & 3\% & $0.24-0.37$ & 21\%\cr
solar with future SNO+3 yrs SK-Gd(low-LMA) & $ 7.0-7.4$ & 3\% & $0.25-0.335$ 
& 14\%\cr
solar with future SNO+3 yrs SK-Gd(high-LMA) & $ 14.5-15.4$ & 3\% & $0.24-0.35$ & 19\%\cr
3 yrs SK-Gd with Kashiwazaki ``down'' & $6.8-7.6$ & 6\%& $0.23-0.40$ & 27\%\cr
7 yrs SK-Gd with {\it only} Shika-2 ``up'' & $7.0-7.3$ & $2\%$ &
$0.28-0.32$ & 6.7\% \cr
\hline
\end{tabular}
\label{klbounds}
\caption
{The range of parameter values allowed at 99\% C.L.
and their corresponding spread.
}
\end{center}
\end{table}

   We begin by reviewing the current status 
of  determination of the solar neutrino 
oscillation parameters. 
We present in Fig. \ref{solnow} the 
regions in the  $\ms-\sss$ plane,
allowed by the global solar neutrino data
(left-hand panel), and
by the combined data from the KamLAND and solar neutrino 
experiments (right-hand panel).
The global solar 
neutrino data used in the analysis include
the total event rates measured in the Homestake \cite{Cl98} 
and SAGE+GALLEX+GNO \cite{SAGE,GALLGNO} (combined) 
experiments, the Super-Kamiokande 44 bin Zenith angle 
spectrum data \cite{SKsol}, 
the 34 bin day-night spectrum data
from the $D_2O$ phase of the SNO experiment \cite{SNO2}
and the charged current (CC), neutral current (NC) and elastic 
scattering (ES) data from the salt phase of SNO \cite{SNO3}
\footnote{We refer the reader to \cite{snonc,salt,snocc} 
for further details of 
the solar neutrino data analysis.}. 
In what concerns the KamLAND results, we use the 
13 binned spectrum data released by the 
\kl collaboration \cite{KamLAND}
\footnote{For details of the statistical 
analysis procedure used for handling 
the \kl spectral data see ref. \cite{kldata}.}.  
Figure \ref{solnow} was obtained using the 
updated standard solar model (BP2004) results
on the solar neutrino 
fluxes and the associated errors \cite{bp2004}. 
The range of allowed values of 
the parameters $\ms$ and $\sss$ are shown 
in Table 1. Also shown  are the 
\% spread, which is defined as,
\be
{\rm spread} = \frac{ a_{max} - a_{min}}
{a_{max} + a_{min}}~,
\label{error}
\ee
%
where $a_{max}~(a_{min})$ are 
the maximal (minimal) allowed
value of the parameter $a$ at 99\% C.L. 

\begin{figure}[t]
\centerline{\psfig{figure=
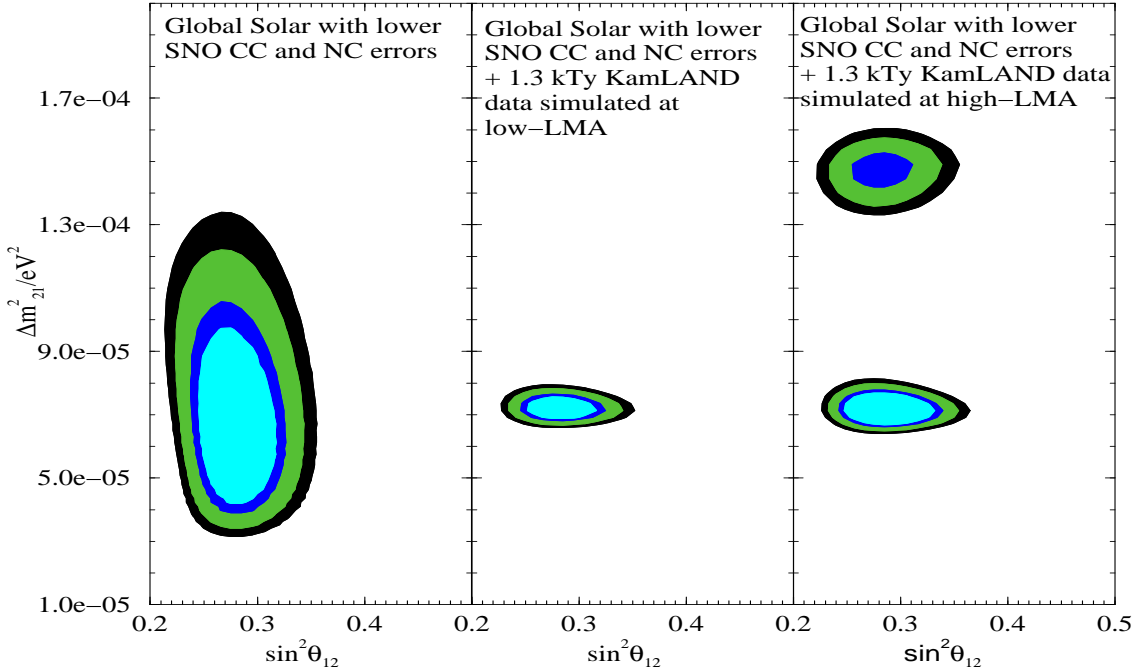,height=9cm,width=15cm}}
\caption{\label{snohe}
The 90\%, 95\%, 99\%, 99.73\% C.L. 
allowed regions in the $\ms-\sss$ plane. The left-panel 
shows the the areas 
obtained in a $\chi^2$ analysis 
of the global solar neutrino data with 
assumed errors of 5\% and 6\% 
on the values of the CC and NC event rates respectively, 
measured in the SNO experiment. The right panel shows the 
allowed region expected if 1.3 kTy \kl data were added 
to the prospective global solar neutrino data (see text for details).
}
 \end{figure}

  The SNO  detector can simultaneously measure the 
CC and NC rates induced by the flux of 
solar $^8B$ neutrinos having energy 
$E_{\nu} \gtap 6.5$ MeV and
$E_{\nu} \gtap 2.2$ MeV, respectively.
While CC rate depends both on 
the $^8B$ flux normalization and the 
solar neutrino survival probability, 
the NC rate is determined, 
in the case of flavor 
oscillation involving only active neutrinos,
solely by the $^8B$ flux normalization. 
This gives SNO the ability to determine the 
average $\nue$ survival probability 
and hence the oscillation 
parameters with a relatively good precision. 
In particular, 
the combination of CC and NC rates in SNO can be effectively 
used to obtain stringent upper limits on the allowed values of
both $\ms$ and $\sss$ \cite{MMSP02,ConchaPGNS00}.
The upper limits on $\ms$ and $\sss$ follow 
from the fact that the ratio 
of the CC and NC event rates
observed in SNO is significantly 
smaller than 0.50 \cite{SNO3}.
The smaller the ratio of 
the CC and NC event rates
observed in SNO, the smaller the 
maximal allowed values of
both $\ms$ and $\sss$ \cite{MMSP02,ConchaPGNS00}.
For values of  $\ms$ and $\sss$
in the low-LMA region, the matter-enhanced 
transitions of $^8B$ neutrinos 
inside the Sun are adiabatic
and the $^8B$ $\nu_e$ 
survival probability,
which affects the CC event rate observed 
in SNO during day-time, 
is  given approximately by
$P_{ee}\approx \sss$.
Thus, improvement in the precision 
with which the CC and NC rates are measured in SNO
would lead to a diminishing of the maximal
allowed values of $\ms$ and $\sss$,
provided the mean values of these 
two observables do not change.
This is illustrated in the left-hand panel of 
Fig. \ref{snohe} where we show the regions of values of 
$\ms$ and $\sss$, which would be allowed by the 
global solar neutrino data
if the experimental errors in the
CC and NC event rates measured by SNO are reduced 
respectively to 5\% and 6\%~
\footnote{The total projected error in the SNO NC 
event rate measurement using Helium  counters
in the  phase III of the experiment 
is expected, according to SNO, to be about 6\% \cite{snotaup}. 
For the CC even rate we assume that the 
statistical error during the phase III would be approximately
the same as in each of the earlier two phases, 
while the systematic error is taken
to be of 4.5\%, i.e., slightly smaller than
the 5\% reported in phases I and II of SNO.  
Thus, we assume that the total error in CC event rate 
measurement from all the three phases 
combined will be about 5\%.},
while the mean values
of the two rates coincide with those
found in the salt phase of the SNO 
experiment \cite{SNO3}.
We note that the maximal allowed values of
both $\ms$ and $\sss$ would be smaller,
while the minimal values essentially 
do not change. Thus, the uncertainties 
in the values of $\ms$ and $\sss$ would be 
reduced with respect to the currently existing ones.  

    Much better precision in the determination of the value of
$\ms$ might be achieved in the \kl 
experiment. In Table 1 we present the 
range of allowed values of this parameter 
and the corresponding spread expected if 
1 kTy data of \kl is combined with 
the current solar neutrino data
(see refs. \cite{talks,prekl} for details of the \kl
prospective data analyses). The expected 
precision  in the determination of $\ms$
from \kl data corresponding to
statistics of 2.6 kTy,  is also shown.  
Since a new reactor power plant 
Shika-2 is planned to start operating 
in March of 2006, we have also included 
the contribution of the flux 
from this new reactor to the 
data collected in \kl after March 2006. The impact of the new
Shika-2 reactor on the \kl sensitivity 
to $\ms$ and $\sss$ was studied in detail 
in \cite{shika}. As was shown in \cite{shika},
with Shika-2 reactor running, 
the precision with which $\sss$ can be determined in 
the \kl  experiment does 
not improve considerably, 
while the high-LMA - low-LMA solution ambiguity 
increases. 

   In the middle (right-hand) panel of Fig. \ref{snohe} we show the expected 
allowed region when we combine a higher precision global 
solar neutrino data, including projected SNO III results, with  
prospective \kl data, simulated at the current low-LMA (high-LMA) 
best-fit point. We expect \kl to have collected 
approximately 1.3 kTy of data by the
end of 2006, when the SNO experiment is foreseen to conclude.
In Fig. \ref{snohe} we present results obtained by combining
the solar neutrino data, including SNO data with 
5\% and 6\% total uncertainties on the 
measured CC and NC event rates, respectively,
with simulated 1.3 kTy \kl data. We again note 
that the allowed range of $\ms$ diminishes 
remarkably with the incorporation of the \kl 
data simulated at the low-LMA best fit point.
With the \kl data simulated at the high-LMA best fit point,
the low-high LMA solution ambiguity ``appears'' at 95\% C.L.
The range of allowed values of $\sss$, 
however, in both cases remains practically unchanged
with the inclusion of the \kl data.

\section{Measurement of the Solar Neutrino Oscillation 
Parameters in the SK-Gd Reactor Experiment}
\label{sens1}

%
%

Super-Kamiokande is the world's 
largest running water Cerenkov detector, 
situated in the Kamioka mine in Japan. 
Next to SK in Kamioka is the \kl detector. 
\kl uses 1 kton of liquid scintillator 
to detect the $\anue$ coming 
from the nuclear reactors. 
SK receives the same reactor antineutrino 
flux as KamLAND. Since SK has ultra pure water as its detector 
material, it cannot efficiently tag the $\anue$ capture on protons.
In particular, the neutron released in $\anue +p\rightarrow e^+ + n$
has to thermalise and then get captured to release  
$\gamma$, which can be then 
detected and the process tagged through 
delayed coincidence. 
The capture cross-section of thermal neutron on free protons 
in extremely small and hence SK is unable to separate the 
reactor $\anue$ capture events from the background.
However, this could change if gadolinium was mixed with the 
SK water, as proposed recently 
in ref. \cite{gdpaper}. 
The capture cross-section of thermal neutron 
on gadolinium is known to be remarkably large.
In addition, neutron capture on Gd produces 8 MeV energy in photons 
making it easier for SK-Gd to detect them.
Thus, SK loaded with Gadolinium could 
be used as a very big reactor anti-neutrino
detector \cite{gdpaper}.
 With its 22.5 kton of ultra pure water, the SK detector has 
about $1.5\times 10^{33}$ free protons as target 
for the antineutrinos coming 
from various reactors in Japan. The \kl detector has only 
$3.46\times 10^{31}$ free target protons \cite{KamLAND}. 
Therefore for the same period of measurements,
SK-Gd detector is expected to 
have about 43 times the statistics 
of the \kl experiment.  
The number of positron events 
in the SK-Gd detector is given by,
\be
N_{SK-Gd} = N_p \int dE_{vis} \int dE_\nu \sigma(E_\nu) R(E_{vis},E_\nu)
\sum_i \frac{S_i(E_\nu)}{4\pi L_i^2}P_i(\anue \rightarrow \anue)
\label{events}
\ee
%
where $E_{vis}$ is the measured {\it visible} energy of the emitted 
positron, when the true visible energy, 
$E_{vis}^T \cong E_{\nu}-0.80$ MeV, 
$E_{\nu}$ being the energy of the incoming $\anue$,
$\sigma(E_\nu)$ is the 
$\anue +p \rightarrow e^+ + n$ reaction
cross-section, 
$S_i(E_\nu)$ denotes the $\anue$ flux 
from the $i$th reactor, 
$L_i$ is the distance between the 
$i$th reactor and Kamioka, 
$ R(E_{vis},E_\nu)$ is the 
energy resolution function of the detector,
$N_P$ are the number of protons in the target, and
$P_i(\anue \rightarrow \anue)$ is the survival probability of the 
$\anue$ coming from the reactor $i$. 
Since the emitted positron from the antineutrino capture on proton 
will behave similarly to the elastically scattered electron 
by solar neutrinos, we use the 
energy resolution function for the solar neutrino detection 
provided by the SK collaboration \cite{SKsol}. 
The SK detector is located 
very close to KamLAND and we take the effective flux $S_i(E_\nu)$ and 
distance $L_i$ for SK-Gd detector to be the same as those for KamLAND. 

\begin{figure}[t]
\centerline{\psfig{figure=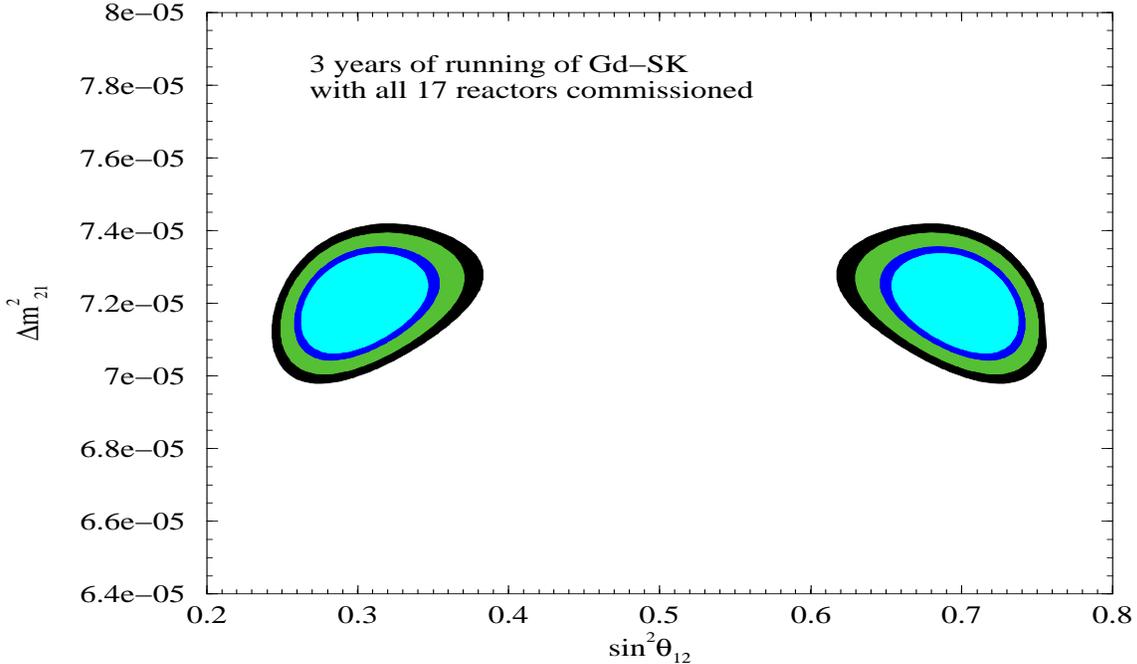,height=9cm,width=15cm}}
\caption{\label{gd3yr} 
The 90\%, 95\%, 99\%, 99.73\% C.L. 
allowed regions in the $\ms-\sss$ plane from an
analysis of prospective data, 
obtained in 3 years of running of the SK-Gd detector  
with all the main 17 reactors running (see text for details).} 
\end{figure}

  We simulate the data expected after 3 years of running of 
the proposed SK-Gd detector at a point in the low-LMA region with 
$\ms=7.2\times 10^{-5}$ eV$^2$ and $\sss = 0.3$; 
for the high-LMA solution the data is simulated in the 
point $\ms=1.5\times 10^{-4}$ eV$^2$ and $\sss = 0.3$ 
\footnote{Unless otherwise stated, all future simulated data in 
this paper correspond to 
$\ms=7.2\times 10^{-5}$ eV$^2$ (low-LMA) or 
$\ms=1.5\times 10^{-4}$ eV$^2$ (high-LMA)
and $\sss = 0.3$.}.
We construct a 18 bin data with a visible energy threshold
of 3 MeV 
\footnote{In order to facilitate 
comparison with the \kl data, 
we present all results in terms 
of the visible energy.}
and with bin width of 0.5 MeV. 
We wish to statistically analyze this prospective data.
We define a $\chi^2$ function given by
\be
\chi^2 =  \sum_{i,j}(N_{i}^{data} - N_{i}^{theory})
(\sigma_{ij}^2)^{-1}(N_{j}^{data} - N_{j}^{theory})~,
\label{chig}
\ee
%
where $N_i^\alpha$ $(\alpha=data,theory)$ 
is the number of events in 
the $i^{\rm th}$ bin, $ \sigma_{ij}^2$ 
is the covariant error matrix 
containing the statistical and systematic 
errors and the sum is over all 
bins. 
The \kl experiment in their first published results 
of the measurement of
reactor $\bar{\nu}_e$ flux  have reported a systematic 
error of 6.48\% \cite{KamLAND}. However, while \kl is a
liquid scintillator detector, SK-Gd would be practically 
a water Cerenkov detector. Therefore the systematics of
the SK-Gd experiment would be 
different from those of the \kl experiment. 
We would expect the SK-Gd systematics to be 
similar to those of the SK solar neutrino experiment, 
since even the energy range of the SK-Gd reactor
$\bar{\nu}_e$ experiment could be approximately the same as that 
of the SK solar neutrino experiment. We therefore assume a 5\%
systematic error in our analysis \cite{SKsol}.

   Figure \ref{gd3yr} shows the allowed regions 
in the $\ms-\sss$ plane, obtained using 3 years of thus simulated 
data in the SK-Gd experiment, when the 
true solution is in the low-LMA region. 
In obtaining this figure we 
have assumed that all 
reactors in Japan, relevant for the
analysis,  including the forthcoming
Shika-2 reactor, would be running with 
their full power. We note that the 
size of the allowed areas 
diminishes  significantly in both $\ms$ and 
$\sss$ compared to the currently allowed regions.
In Table 1 we show the uncertainties in $\ms$ and $\sss$ 
after three years of running of SK-Gd alone. We also give the 
expected uncertainties if data from  
5 years of running 
of the SK-Gd detector are used in the analysis.

\begin{figure}[t]
\centerline{\psfig{figure=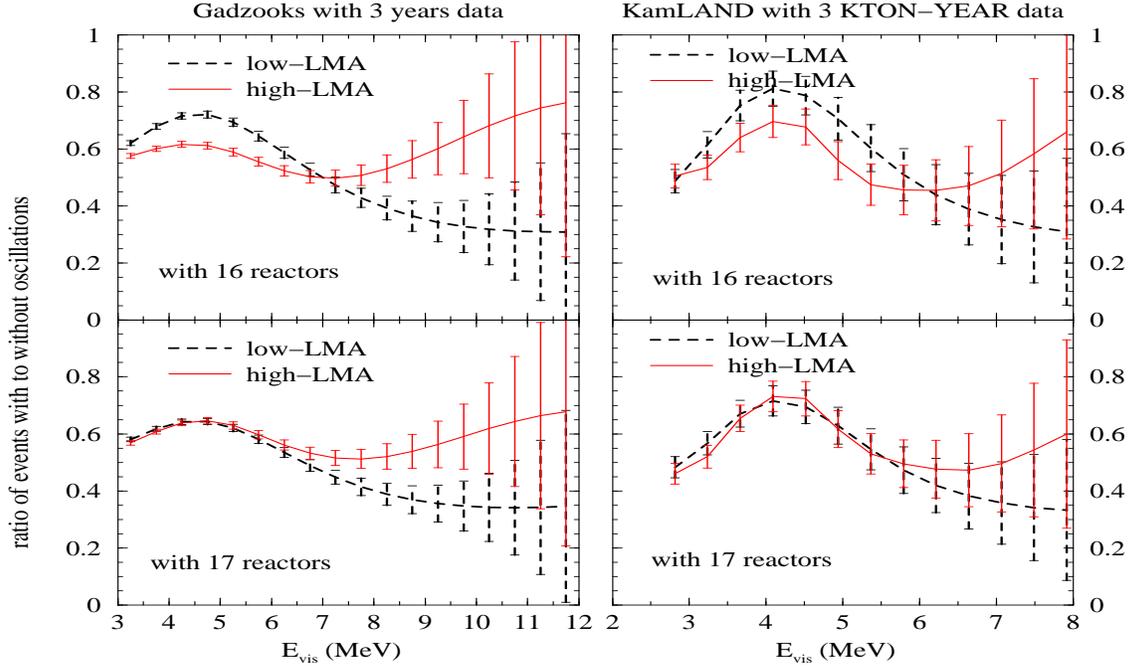,height=9cm,width=15cm}}
\caption{\label{spmgdkl}
The ratio of the numbers of the prospective $e^+-$ events 
at the SK-Gd and KamLAND experiments
in the cases of oscillations and of absence of oscillations,
with their expected $1\sigma$ errorbars.  
The errorbars for the SK-Gd detector correspond to 
3 years of data, while those for \kl are for a total
statistics of 3 kTy. The upper panels 
are obtained assuming that the current 
16 main reactors in Japan are operative. The bottom panels 
correspond to the case of the 
Shika-2 reactor operating along with the current 
16 main reactors. The dashed (solid) lines 
correspond to the low-LMA (high-LMA) 
solution.
}
\end{figure}

  As Fig. \ref{gd3yr} shows, the spurious high-LMA solution 
would be completely ruled out by the SK-Gd data, 
if low-LMA is the true solution.
We have checked that SK-Gd can rule out 
the ``wrong'' solution with 
about $0.4-0.5$ years of data.
This is in sharp contrast with what we have 
obtained for KamLAND, where it is hard to 
resolve the low-LMA -- 
high-LMA ambiguity even with 2.6 kTy of data 
once the Shika-2 reactor is switched on. 
To facilitate the understanding of this result 
we present in Fig. \ref{spmgdkl} the 
spectral distortions predicted to be observed 
in the SK-Gd (left panels) 
and in KamLAND (right panels) experiments. 
The upper panels show the 
ratio of the events (or event rates) 
in the cases of oscillations and of
absence of oscillations,
with the current main 16 reactors running, 
while the lower panels are for 
the case where the Shika-2 reactor 
is operating along with the other 16 reactors.
The dashed black lines are the expected spectra for 
the low-LMA solution, while the red solid lines are the spectra 
for the high-LMA solution. The errorbars correspond to the 
1$\sigma$ statistical errors for 3 years 
of running of the SK-Gd detector and for
3 kTy ($\sim 7$ years) of running of the KamLAND
detector. The bin sizes used are
0.5 MeV for SK-Gd and 0.425 MeV for KamLAND. 
The threshold for KamLAND is taken as $E_{vis}^{th}=2.6$ MeV 
\cite{KamLAND}, while for SK-Gd we use a threshold of
$E_{vis}^{th}=3.0$ MeV. For the SK-Gd detector 
the observed spectrum  extends to nearly 12 MeV 
due to the worse energy resolution of SK.
Clearly,  in the 16 reactor case 
the SK-Gd experiment can distinguish between the 
low-LMA and  high-LMA spectra.  
The \kl experiment could also distinguish 
between the two solutions with a 3 kTy statistics 
(see also \cite{th12,shika,prekl}) 
\footnote{Let us note that using combined data from the \kl detector,
corresponding to statistics of 1kTy, and from the solar neutrino
experiments, can permit to rule out the high-LMA solution
 \cite{salt,shika}.}.
The differences between the $e^+-$spectra 
in the cases of the low-LMA and the high-LMA solutions, 
measured in SK-Gd and KamLAND experiments,
diminish once the 
Shika-2 reactor is switched on. 
The low-LMA and high-LMA solution spectra 
measured with the KamLAND detector
would be almost overlapping, as can be
seen in the lower right-hand panel. 
This results in the reappearance 
of the ``wrong'' high-LMA solution, 
as stressed in \cite{shika}. Even for the
SK-Gd detector, the difference between 
the spectra corresponding to the  two solutions reduces 
considerably, especially in the statistically 
most relevant bins.
However, there are still bins at intermediate energies at which the 
two solutions can be easily distinguished. 
This is why the SK-Gd experiment could still 
resolve the ambiguity between the low-LMA and high-LMA
solutions with Shika-2 reactor operating, 
while \kl cannot.

   The mixing angle is most precisely determined 
when the baseline corresponds to a 
minimum in the survival probability 
(SPMIN) in the statistically most 
relevant part of the positron 
energy spectrum \cite{th12}. 
It was pointed out in \cite{shika} that 
for the low-LMA solution
the baseline from the Shika-2 reactor to Kamioka 
is such that the SPMIN for the Shika-2 flux occurs
at $E_{\rm vis} \cong 4 MeV$.
The Kashiwazaki-Kariwa power plant, which is the most 
powerful reactor complex in the world, is at a distance 
of about 160 km from Kamioka. Hence, the most significant part of
the flux of $\bar{\nu}_e$ from this most important source
falls at Kamioka in the region of a survival probability maximum
(SPMAX). 

   The effective flux 
at Kamioka in absence of oscillations 
is defined as
\be
\Phi_{KL}^i = \frac{P_i}{4\pi L_i}.
\ee
%
For the Kashiwazaki complex 
it corresponds to
$\Phi_{KL}\approx 7.3 \mu W/cm^2$,
while for the Shika-2 reactor 
it is substantially smaller, $\Phi_{KL}\approx 4.1 \mu W/cm^2$.
Thus, the contributions to the 
signals observed in SK-Gd and
\kl due to the Kashiwazaki $\anue$ flux
will be considerably larger than that
due to the Shika-2 flux,
when the contributions from all the 
reactors are combined. Therefore 
the resultant $e^+-$spectrum due to the cumulative 
$\bar{\nu}_e$ flux at Kamioka, produced by 
all relevant reactors in Japan,
still exhibits the effect of the
$\bar{\nu}_e$ survival probability maximum.
This is seen also in Fig.  \ref{spmgdkl}.

\begin{figure}[t]
\centerline{\psfig{figure=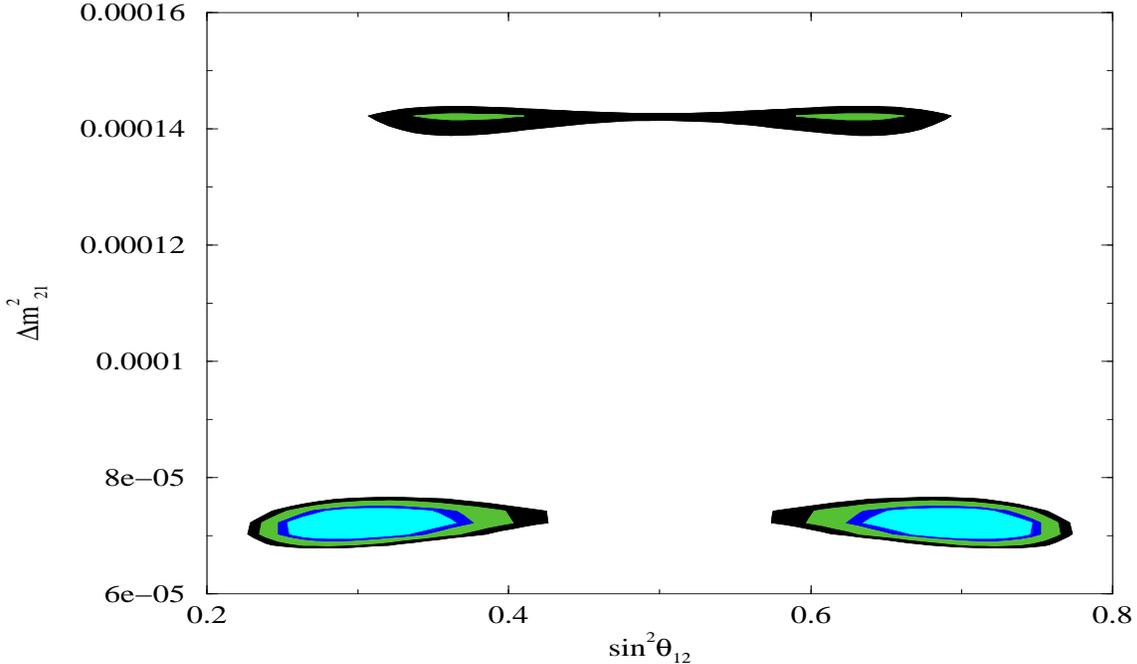,height=9cm,width=15cm}}
\caption{\label{gd3yr-kashi0}
Allowed regions in the $\ms-\sss$ plane from an analysis
of prospective data from the SK-Gd detector collected over a period 
of 3 years, with the Kashiwazaki-Kariwa complex 
switched off and all other reactors, 
including the Shika-2 one, 
running.}
\end{figure}

\begin{figure}[t]
\centerline{\psfig{figure=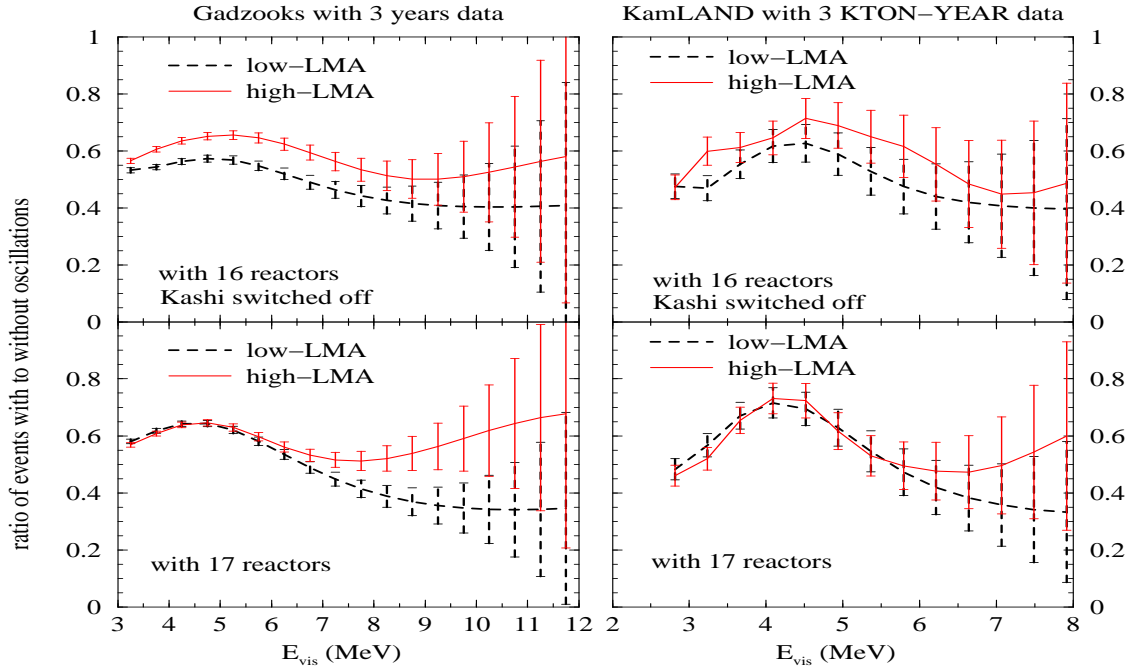,height=9cm,width=15cm}}
\caption{\label{spmgdkl-kashi0}
The ratio of the number of $e^+-$events in the cases of oscillations
and of absence of oscillations 
with their $1\sigma$ errorbars, expected 
at SK-Gd (3 years of data) and KamLAND (3 kTy of data) experiments.  
The Shika-2 reactor is supposed to be operating.
The upper panels are obtained assuming that the Kashiwazaki-Kariwa power 
plant is completely switched off, while the
lower panels correspond to all the 17 reactors, 
including Kashiwazaki-Kariwa and Shika-2, working.
The dashed (solid) lines correspond to the low-LMA (high-LMA) 
solution.
}
\end{figure}

   We therefore consider a fictitious scenario in which
the Kashiwazaki complex is completely switched off, 
while the Shika-2 is operating. 
In figure \ref{gd3yr-kashi0} we show 
the allowed regions obtained using 
the data collected by the
SK-Gd detector after 3 years of operation 
with Kashiwazaki reactor complex switched off, 
but with the Shika-2 reactor working. 
We show the uncertainty on 
$\ms$ and $\sss$, determined with this set up, in Table 1. 
We note that instead of diminishing, 
the uncertainty in $\sss$ increases.
Even the spurious high-LMA solution reappears in this case.

  The predicted
spectral distortions 
for the low-LMA and high-LMA solutions
expected to be observed
in SK-Gd and KamLAND, when the Kashiwazaki-Kariwa power 
plant does not operate, while the Shika-2 one 
is running at its full power,
are shown in the upper panels in Fig \ref{spmgdkl-kashi0}. 
For comparison we also show the spectrum 
expected with all the main 17 reactors
operating at their full strength.
A comparison of the upper and 
lower left-hand panels indicates that with the 
Kashiwazaki complex switched off, the 
distortions in the expected spectrum for the low-LMA solution 
become smaller. We note that even though 
the spectra corresponding 
to the high-LMA  and low-LMA solutions
and measured with the SK-Gd detector 
are less overlapping, the difference in the 
spectral {\it shape} for the two solutions
diminishes when the Kashiwazaki plant 
is not operating. Thus, it could be 
possible with a 
value of $\ms$ in the high-LMA zone, to 
describe the data simulated at the 
``true'' low-LMA solution point.
However, in this case  
a somewhat larger value of 
$\sss$ is required in order
to explain the larger reduction of the
event rate corresponding to
the value of $\ms$ of  
the ``true'' low-LMA solution.
This is what we get in 
Fig.  \ref{gd3yr-kashi0}, where the 
spurious high-LMA solution 
gets allowed with a higher value of $\sss$.

 To stress the importance of having an experimental set up 
in which there are no cancellations between SPMIN and SPMAX 
due reactors at very different distances, we consider a 
hypothetical situation where only one reactor is operating.
As a concrete example we consider a scenario where only the Shika-2
reactor is operating. The flux from this reactor 
would produce a SPMIN in the $e^+-$spectrum detected at SK-Gd
if low-LMA was the correct solution.
We show in Table 1
the range of allowed parameter values and the corresponding 
spread at 99\% C.L. obtained with 7 years 
of data recorded in SK-Gd, when {\it only} Shika-2 is 
operating. The improvement in the precision of $\sss$ 
measurement is seen to be remarkable. 

\section{Combined Impact of Solar and SK-Gd Data}
\label{solgdsec}

\begin{figure}[t]
\centerline{\psfig{figure=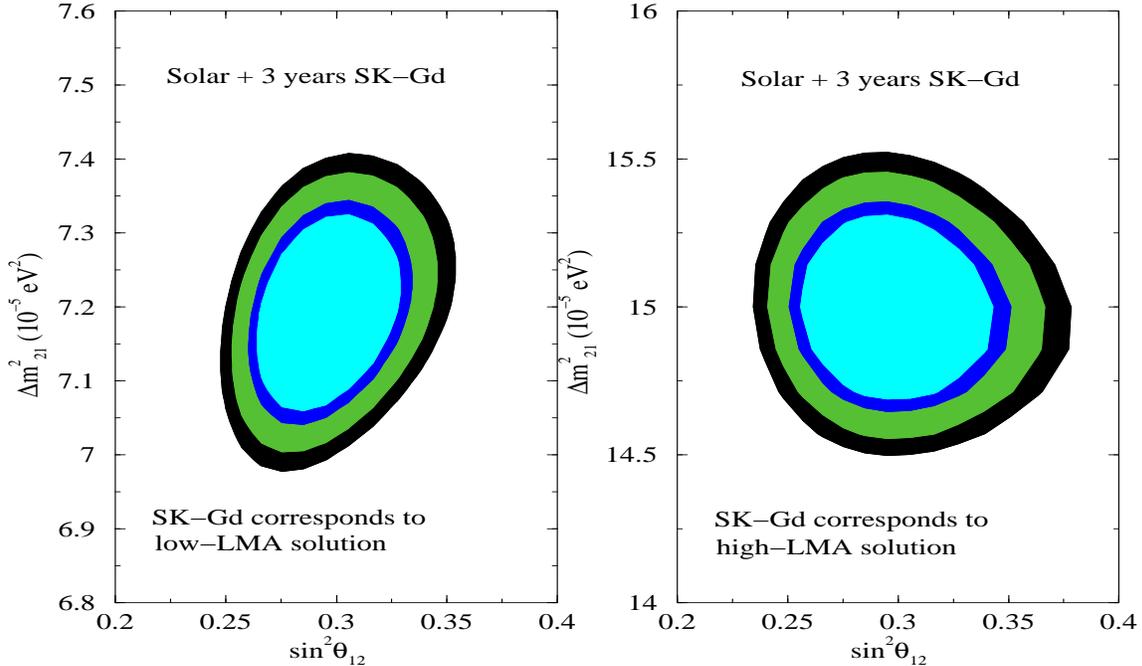,height=9cm,width=15cm}}
\caption{\label{solgd}
The 90\%, 95\%, 99\%, 99.73\% C.L. regions
in the $\ms-\sss$ plane,
allowed by the combined data from the solar neutrino experiments and 
from the SK-Gd detector after
3 years of running of the latter. The left-hand panel shows 
the allowed regions 
obtained in the case the SK-Gd data corresponded to the low-LMA
solution, 
while  the right-hand panel shows the allowed regions 
if high-LMA was the true solution.
}
\end{figure}
\begin{figure}[t]
\centerline{\psfig{figure=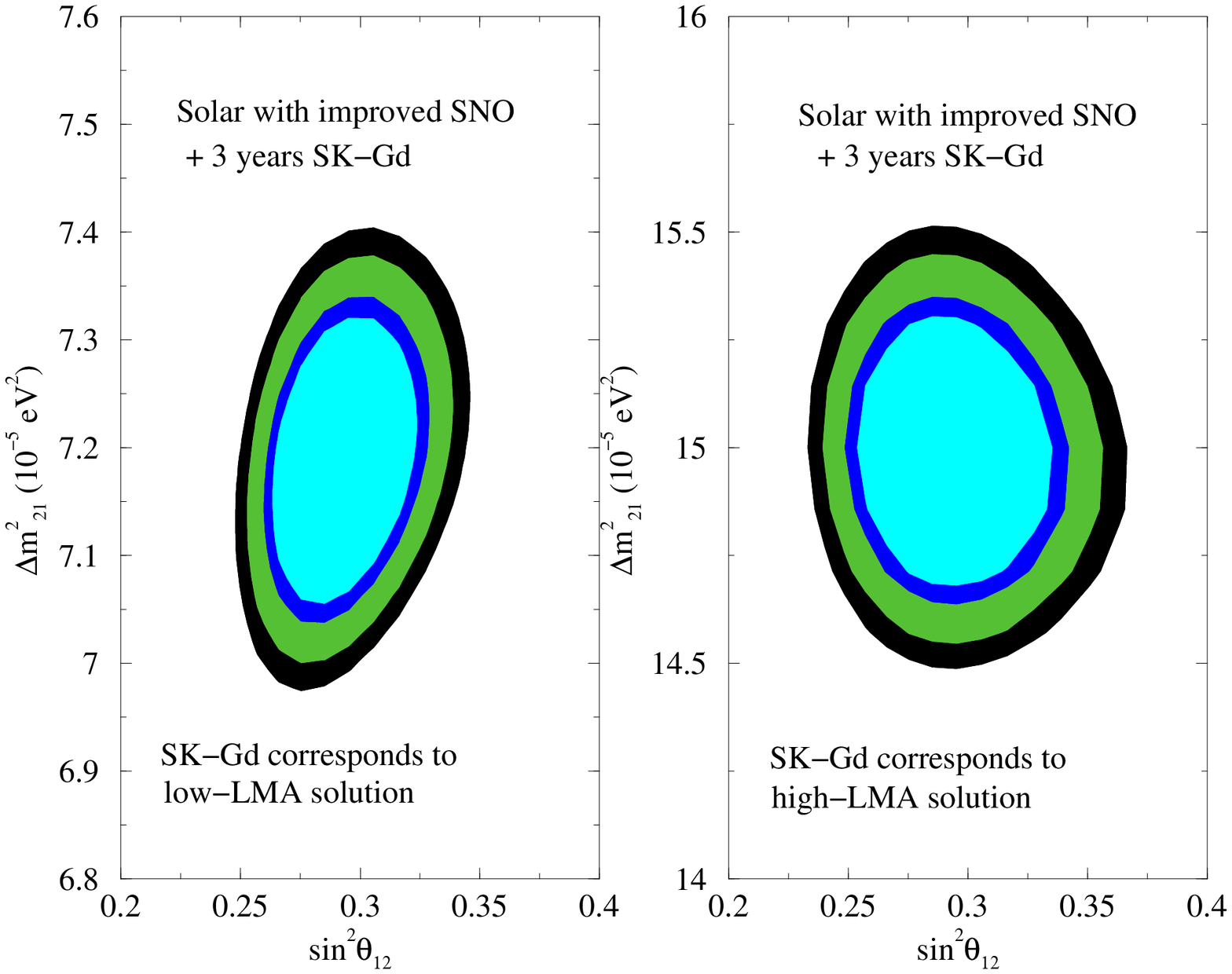,height=9cm,width=15cm}}
\caption{\label{sol3gd}
The same as in Fig. \ref{solgd}, but with errors on the
CC and NC event rates measured at SNO reduced 
to 5\% and 6\%, respectively.}
\end{figure}

In Fig.  \ref{solgd} we present the allowed regions obtained by a 
combined analysis of the current global solar neutrino data and 
prospective SK-Gd data, expected after 3 years of running of 
the proposed SK-Gd experiment.
The left-hand panel shows the  
case of low-LMA being the true solution 
of the solar neutrino 
problem; we simulated the SK-Gd data 
at $\ms=7.2\times 10^{-5}$ eV$^2$ and $\sss=0.3$. 
The right-hand panel 
illustrates the scenario we would witness 
if against the current 
trend, the next set of data from the KamLAND experiment
would favor the high-LMA solution. The 
SK-Gd data for the right-hand panel is simulated at 
$\ms=1.5\times 10^{-4}$ eV$^2$ and $\sss=0.3$. 

In the case of the low-LMA 
assumed to be the true solution 
of the solar neutrino problem,
the inclusion of SK-Gd data 
from 3 years of measurements 
in the global solar neutrino 
analysis reinforces the low-LMA solution. The 
best-fit and the range of allowed values of $\ms$ are determined 
almost solely by the SK-Gd data. The best-fit and the range of allowed 
values of $\sss$ depend on both the solar and the SK-Gd data, 
though again due to its enormous statistics the SK-Gd data have 
larger impact on the $\sss$ determination.

  If, however, the next \kl results do conform to the high-LMA solution, 
we would encounter a conflicting trend where 
the description of the 
solar neutrino data and of the \kl data would 
require different values of $\ms$. 
Such a situation might warrant 
an  explanation which includes 
new phenomenon along with matter induced 
oscillations to be responsible for the solar neutrino deficit 
(see,  e.g.,  \cite{Guzzo}).  
The data collected by SK-Gd experiment in 3 years, when combined with 
the global solar neutrino data, would lead to the
allowed regions in the $\ms -\sss$ plane, 
shown in the right-hand panel in Fig.  \ref{solgd}.
We note that if the SK-Gd data conformed to a point in the high-LMA 
zone, the low-LMA solution which is favored by the solar neutrino 
data, would be  completely ruled out. 
This reflects the statistical power 
of the SK-Gd experiment. In a similar analysis with 
prospective \kl data only (and no SK-Gd data)
corresponding to the high-LMA solution, 
we would get allowed zones in 
both the low-LMA  and high-LMA regions (see, e.g., ref. \cite{salt}).

In Table 1 we present the range of allowed $\ms$ and
$\sss$ values, determined in a combined analysis of the solar neutrino
and  SK-Gd data. In the case of 
the true value of $\ms$ lying in the low-LMA region, 
the lower limit on $\ms$ is less stringent in  comparison 
with the lower limit one obtains from the 3 year data of SK-Gd only. This 
is due to the solar neutrino data favoring values 
of $\ms \approx 6\times 10^{-5}$ eV$^2$. 
The upper limit on $\ms$ is determined solely by the SK-Gd data.
The maximal allowed value of $\sss$ diminishes 
with the inclusion of the solar neutrino
data in the analysis together with the SK-Gd data. Correspondingly, 
the spread reduces from 19\% to 15\%. 
The lower limit on $\sss$ is determined  
by the SK-Gd data.

 If the SK-Gd data  is simulated at a point in the high-LMA
region, $\ms$ has a spread of 3\% and $\sss$ has a spread of 
12\%. We note that the spread in $\sss$ is larger for values of 
$\ms$ in the high-LMA region than for $\ms$ lying in the 
low-LMA region. The 
accuracy in the $\ms$ determination 
is somewhat worse 
in the case of the high-LMA solution compared
to the case of the low-LMA solution 
because there is less spectral 
distortion for the $\ms$ values in the high-LMA zone, 
as can be seen in Fig.  \ref{spmgdkl}.
For a high precision determination of $\sss$, 
conditions as close to  SPMIN as possible 
are required \cite{th12,shika}. We find 
that the precision in $\sss$ is worse for the high-LMA solution 
since the spectrum has a smaller ``dip''
even for the higher energy bins, as  compared to
the ``dip'' in the spectrum in the case of the 
low-LMA solution (this is valid even if the Shika-2 
reactor is operating, see the lower left-hand 
panel in Fig. \ref{spmgdkl}).

In Fig. \ref{sol3gd} we present the the allowed regions obtained by a 
combined analysis of the global solar neutrino data with reduced 
errors for the SNO CC (5\%) and SNO NC (6\%) and 
prospective SK-Gd data, expected after 3 years of running of 
the proposed SK-Gd experiment. In Table 1 we present the 
corresponding allowed range for $\ms$ and $\sss$ and the spread 
at 99\% C.L.

\section{Conclusions}
\label{concl}

  In this paper we have studied the 
prospects of high precision determination of 
the solar neutrino oscillation parameters,
$\ms$ and $\sss$,
by investigating the reactor  $\bar{\nu}_e$ 
oscillations with the recently proposed 
gadolinium loaded Super-Kamiokande
(SK-Gd) experiment \cite{gdpaper}. 
Doping the  SK detector with Gadolinium 
by dissolving 0.2\% gadolinium trichloride in the water
\cite{gdpaper} would make it possible to 
detect the neutrons, released 
in the $\anue$ capture on protons, with a 
relatively high efficiency. 
This would allow the SK experiment, 
in particular, to detect the reactor $\bar{\nu}_e$ 
coming from the numerous powerful nuclear 
reactors located in Japan, thus transforming SK
into the largest long baseline reactor 
$\bar{\nu}_e$ detector with 
an event rate approximately 
43 times higher than 
that observed in the KamLAND experiment.

 Working in the framework of two-neutrino mixing,
we have presented first an overview of the 
currently allowed ranges of values of the
solar neutrino oscillation parameters 
$\ms$ and $\sss$ by the existing global solar 
neutrino and KamLAND data (Fig. 1 and Table 1),
as well as the regions in the
parameter space 
which are expected to be allowed 
after taking into account 
future higher statistics and 
lower systematic data from SNO and from KamLAND
experiments (Fig. 2). 
The SK-Gd detector
is considered  in Section \ref{sens1},
where we discuss the assumptions 
made to simulate the data in this 
proposed modification of the SK detector, 
as well as the procedure used to statistically
analyze the simulated data. 

   The results of our analysis show
that the SK-Gd experiment 
has a remarkable potential 
in reducing the uncertainties in the values of 
$\ms$ and $\sss$ (Figs. 3 and 7 and Table 1).
Combining the SK-Gd data, taken over a period of 3 years,
with the data from the solar neutrino experiments, 
will allow to determine 
$\ms$ with a 3\% error at 99\% C.L., the error
being by a factor of 2 smaller if one uses
only the SK-Gd data.
With only $\sim 0.5$ years of data, 
the SK-Gd experiment
would rule out completely the spurious 
high-LMA solution if low-LMA is the true solution.
In contrast, the high-LMA solution could be ruled out 
if one uses the data from the solar neutrino
experiments and data from the \kl detector,
corresponding to 1kTy of statistics. 
If the future \kl data would conform to a point 
in the high-LMA region, both the high-LMA and the low-LMA 
solutions would be allowed at 90\% C.L. \cite{salt}.
In this case the high-LMA -- low-LMA 
solution ambiguity can be resolved
completely by the SK-Gd experiment.
In what regards the solar neutrino mixing angle,
using the data from the solar neutrino experiments
and the SK-Gd 3 year data
will allow to determine 
$\sss$ with a $\sim 15\%$ error at 99\% C.L.
One would have the same uncertainty 
in $\sss$ using data only from the SK-Gd 
experiment, corresponding 
to 5 years of measurements.
Thus, the SK-Gd experiment could provide one of the most
precise (if not the most precise) determination of the
solar neutrino oscillation parameters
$\ms$ and $\sss$.

\section*{Addendum after the 766.3 kTy results from KamLAND}

\begin{figure}[t]
\centerline{\psfig{figure=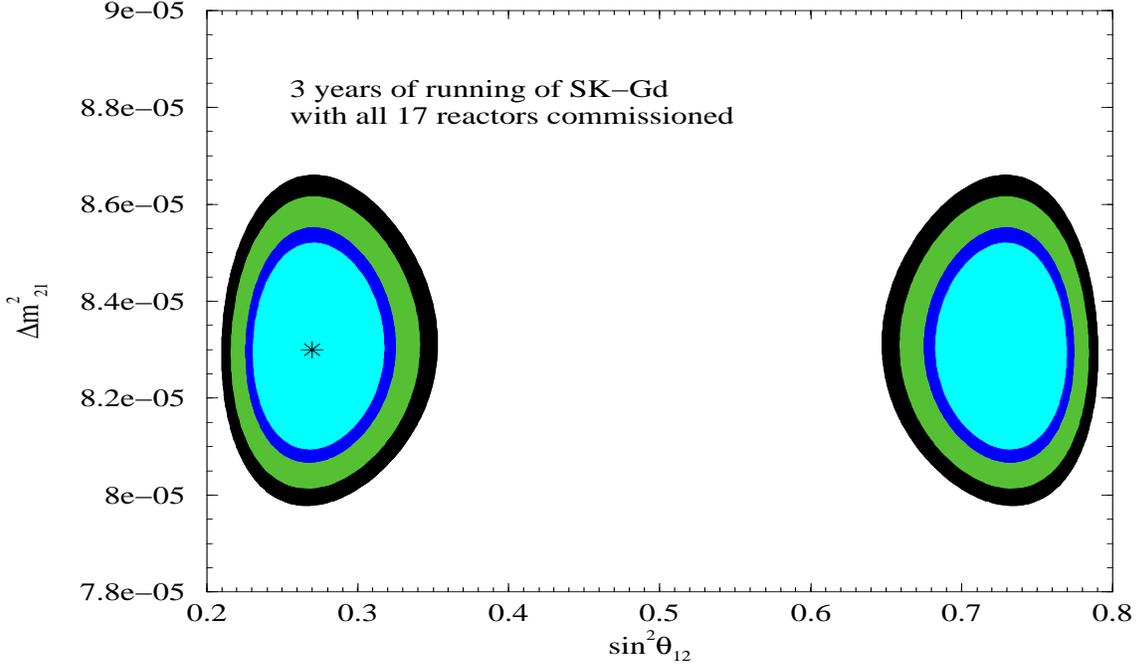,height=9cm,width=15cm}}
\caption{\label{gd8.33yr}
Same as Fig. \ref{gd3yr} but with the SK-Gd data simulated at the 
new global best-fit point, $\ms=8.3\times 10^{-5}$ eV$^2$ 
and $\sss=0.27$, marked on the figure by a star.
}
\end{figure}

\begin{figure}
\centerline{\psfig{figure=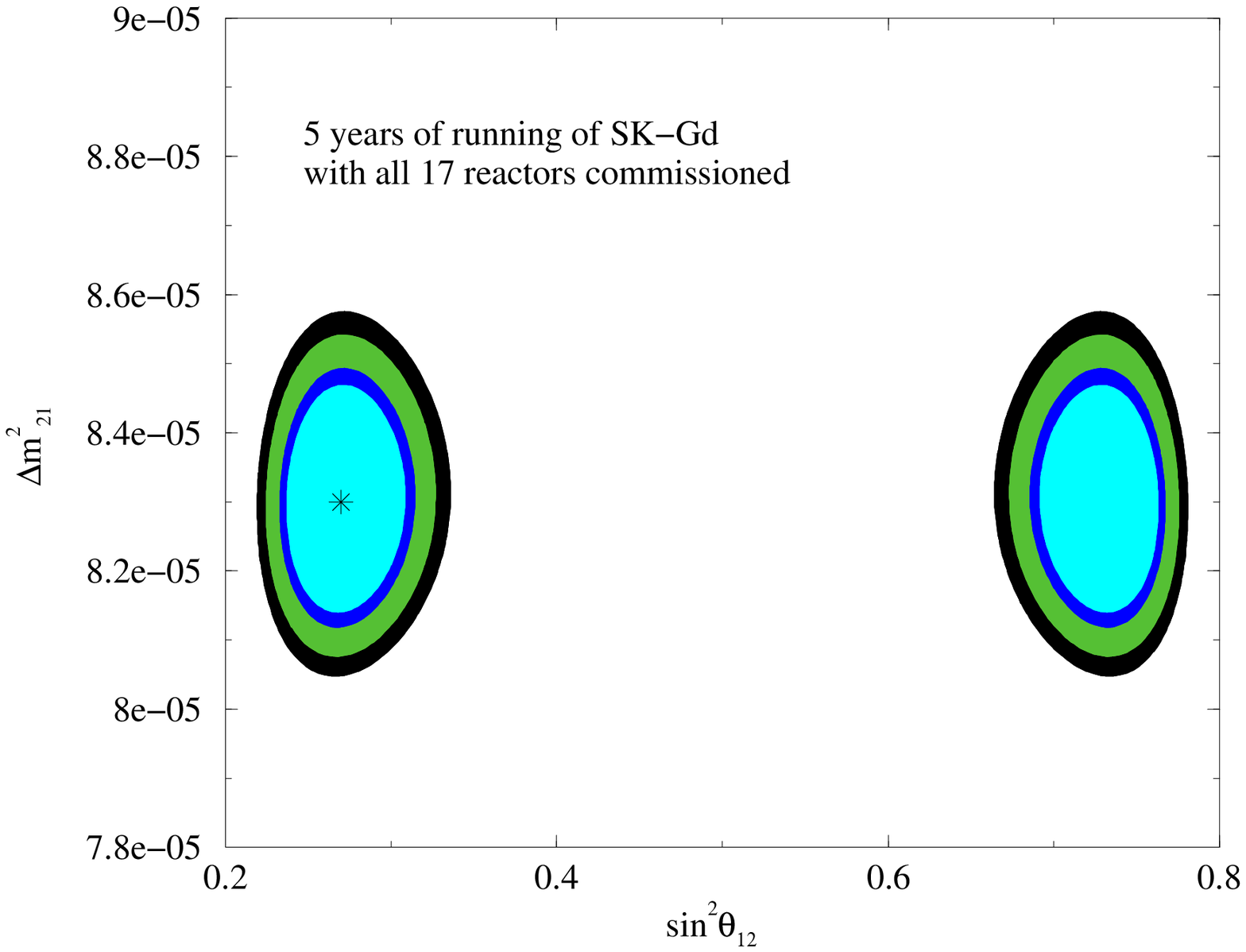,height=9cm,width=15cm}}
\caption{\label{gd8.35yr}
Same as Fig. \ref{gd8.33yr} but for 5 years statistics.
}
\end{figure}

The KamLAND collaboration 
has published data corresponding to a statistics of 766.3 kTy 
\cite{kl766}. The new results from KamLAND have corroborated their 
first results from 2002 and have for the first time provided
unambigous evidence for distortion in the measured reactor 
antineutrino spectrum. These new results have reinforced the low-LMA 
solution of the solar neutrino problem and the combined solar and
\kl data now rule out the high-LMA solution at the $4\sigma$ level 
\cite{kl2our}. The best-fit values of the solar neutrino oscillation
parameters obtained from the combined 
analysis of solar and \kl data are $\ms=8.3\times 10^{-5}$ eV$^2$ 
and $\sss=0.27$ and their corresponding 
99\% C.L. allowed range of values are given 
as
\bea
\ms &=& (7.3-9.4)\times 10^{-5} {\rm eV^2};~ ~ {\rm spread=13\%} 
\\
\sss &=& (0.22-0.36);~~ {\rm spread=24\%} 
\eea

We present in this section the allowed regions in the $\ms-\sss$ 
parameter space expected from the Sk-Gd experiment when 
the ``data'' is simulated at the new global best-fit point 
$\ms=8.3\times 10^{-5}$ eV$^2$ 
and $\sss=0.27$. The Fig. \ref{gd8.33yr} shows the 
allowed regions for 
3 years statistics in SK-Gd.
The corresponding allowed range of $\ms$ and $\sss$ 
and their spread, expected at 99\% C.L. is given as
\bea
\ms &=& (8.01-8.61)\times 10^{-5} {\rm eV^2};~ ~ {\rm spread=3.6\%} 
\\
\sss &=& (0.22-0.34);~~ {\rm spread=21\%} 
\eea

Fig. \ref{gd8.35yr} shows the  
allowed regions for 
5 years statistics in SK-Gd with the spectrum simulated at the 
new global best-fit. The 99\% C.L. allowed range and spread in 
$\ms$ and $\sss$ are given as
\bea
\ms &=& (8.07-8.53)\times 10^{-5} {\rm eV^2};~ ~ {\rm spread=2.8\%} 
\\
\sss &=& (0.22-0.32);~~ {\rm spread=18\%} 
\eea
We note that the precision on $\ms$ and $\sss$ measurement in 
SK-Gd for a given statistics remains roughly the same for the 
old and new global best-fit points.


\section*{Acknowledgements}
The authors would like to thank M. Vagins for useful 
correspondence. The results on the solar neutrino data analysis 
presented in this work were obtained in collaboration with 
A. Bandyopadhyay and S. Goswami, and the authors wish to thank them. 
S.T.P. would like to thank Prof. T. Kugo,
Prof. M. Nojiri and the other members of the
Yukawa Institute 
for Theoretical Physics (YITP), Kyoto, Japan,
where part of the work 
on present study was done,
for the kind hospitality extended to him. 
This work was supported in part 
by the Italian INFN under the program 
``Fisica Astroparticellare''.


\end{document}